\documentclass[pra,a4paper,twocolumn,showpacs]{revtex4}
\usepackage{amssymb}
\usepackage{amsmath}
\begin{document}
\title{Exchange Monte Carlo for Molecular Simulations with
Monoelectronic Hamiltonians}
\author{F. Calvo and F. Spiegelman}
\affiliation{Laboratoire de Physique Quantique, IRSAMC, Universit\'e Paul
Sabatier, 118 Route de Narbonne, F31062 Toulouse Cedex}
\begin{abstract}
We introduce a general Monte Carlo scheme for achieving atomistic simulations
with monoelectronic Hamiltonians including the thermalization of both nuclear
and electronic degrees of freedom. The kinetic Monte Carlo algorithm is used
to obtain the exact occupation numbers of the electronic levels at canonical
equilibrium, and comparison is made with Fermi-Dirac statistics in infinite
and finite systems. The effects of a nonzero electronic temperature on the
thermodynamic properties of liquid silver and sodium clusters are presented.
\end{abstract}
\maketitle

In the recent years, the physics of materials and complex systems has undergone
awesome development in the field of molecular simulation \cite{molsim}.
Significant achievements include {\em ab initio}, Car-Parrinello molecular
dynamics (MD) \cite{cpmd}, linear-scaling \cite{linescale}, or
progresses in ergodic techniques such as exchange Monte Carlo (EMC)
\cite{ptmc}, also known as parallel tempering, or the Wang-Landau
algorithm \cite{wanglandau}.
Account of electronic structure is often made either implicitely with
empirical potentials or more explicitely through one-electron methods, either
in the framework of density-functional theory (DFT) or using tight-binding (TB)
approximations. In the latter cases, the electronic ground state is described
via integer occupation numbers. This is usually adequate for nonmetal systems.
However the electronic states of metals can cross the Fermi
surface, therefore integer occupations are not appropriate for continuous
dynamics. Fractional occupation numbers must also be introduced for
insulators or semiconductors, provided that the temperature is high enough for
the lowest excited states to be populated.

Building upon a seminal paper by
Mermin \cite{mermin} who extended the Hohenberg-Kohn theorem of
DFT to nonzero electronic temperatures, several
authors proposed to combine DFT and fractional occupation
numbers \cite{weinert,alavi,springborg}.
Most of these works were devoted to
metals, and they tried to find suitable forms of the occupation
laws for numerical stability purposes and a better sampling in the Brillouin
zone. In particular, it was proposed to
replace the Fermi-Dirac (FD) function with other expressions \cite{springborg}.

Even when keeping a FD distribution, the best choice for electronic temperature
was shown not to be necessarily related with the nuclear vibrational
temperature \cite{mehl}. Additionally, and strictly speaking, the FD
function holds for an infinite system, but should only be considered as an
approximation when treating a small molecular system such as a metal cluster.

The goal of this Letter is to show how the true canonical equilibrium of both
ionic and electronic degrees of freedom can be simulated through MC
methods, for small or large sizes. A specific interest in choosing MC methods
over MD simulations is their greater flexibility and wider range of
application. They are very convenient for discrete systems, and they can be
adapted using statistical biases to
accelerate convergence. They also offer a straightforward way to sample
grand-canonical ensembles, which is more difficult with MD \cite{molsim}.
As seen below, the Monte Carlo method is well suited to the
problem of electronic thermalization, especially for finite systems.

We consider the general class of materials modelled by monoelectronic
Hamiltonians. At any given nuclear configuration ${\bf R}$, the total energy
$E$ depends on the occupation numbers $\{n_i\}$. In the Kohn-Sham
formalism, these numbers appear explicitely in the expression of the density,
hence in the energy. In TB models, the band contribution is the weighted sum
of the one-electron energies $\{\varepsilon_i\}$: $E({\bf R}) = \sum_i n_i
\varepsilon_i({\bf R})$.
The first, most simple MC algorithm consists in treating the nuclear
$({\bf R})$ and electronic $({\bf N})$ degrees of freedom on the same footing,
by performing random moves on the generalized coordinates ${\bf Q}= ({\bf
R}, {\bf N})$. Here ${\bf N}= \{ n_i\}$ is the set of instantaneous integer
occupation states, which evolve during the simulation accordingly with the
level statistics. In the Metropolis scheme, a change from ${\bf
Q}_{\rm old}$ to ${\bf Q}_{\rm new}$ due to a change in either ${\bf R}$ or
${\bf N}$ is accepted with probability ${\rm acc}({\bf Q}_{\rm old} \to
{\bf Q}_{\rm new}) = \min[1,\exp(-\beta \Delta E)]$ where $\Delta E =
E({\bf Q}_{\rm new})-E({\bf Q}_{\rm old})$.
The MC moves involving ${\bf N}$ must keep constant the total number ${\cal
N}$ of occupied states. The only moves of this kind that we consider are
single exchanges between occupied and
unoccupied, {\em neighboring}\/ levels. For a given ionic geometry ${\bf
R}$, this algorithm clearly converges to the electronic canonical distribution
in the ergodic limit.

The local moves involving ${\bf R}$ and ${\bf N}$ are now implemented in
the framework of generalized ensembles. In our case, we use exchange
Monte Carlo by performing simultaneous simulations at
various temperatures. With some probability $p$, exchange moves between
${\bf Q}_i = ({\bf R}_i,{\bf N}_i)$ and ${\bf Q}_j = ({\bf R}_j,{\bf N}_j)$ at
the inverse temperatures $\beta_i$ and $\beta_j$, respectively, are attempted.
They are accepted with the probability
\begin{equation}
{\rm acc}({\bf Q}_i \rightleftharpoons {\bf Q}_j) = \min[1,\exp(\Delta \beta
\Delta E)].
\label{eq:ptmc1}
\end{equation}
In this equation, $\Delta \beta = \beta_i - \beta_j$ and $\Delta E = E({\bf
Q}_i)-E({\bf Q}_j)$. With probability $1-p$, the usual local moves are
attempted for all trajectories.

\begin{figure}[htb]
\vbox to 6cm{
\includegraphics{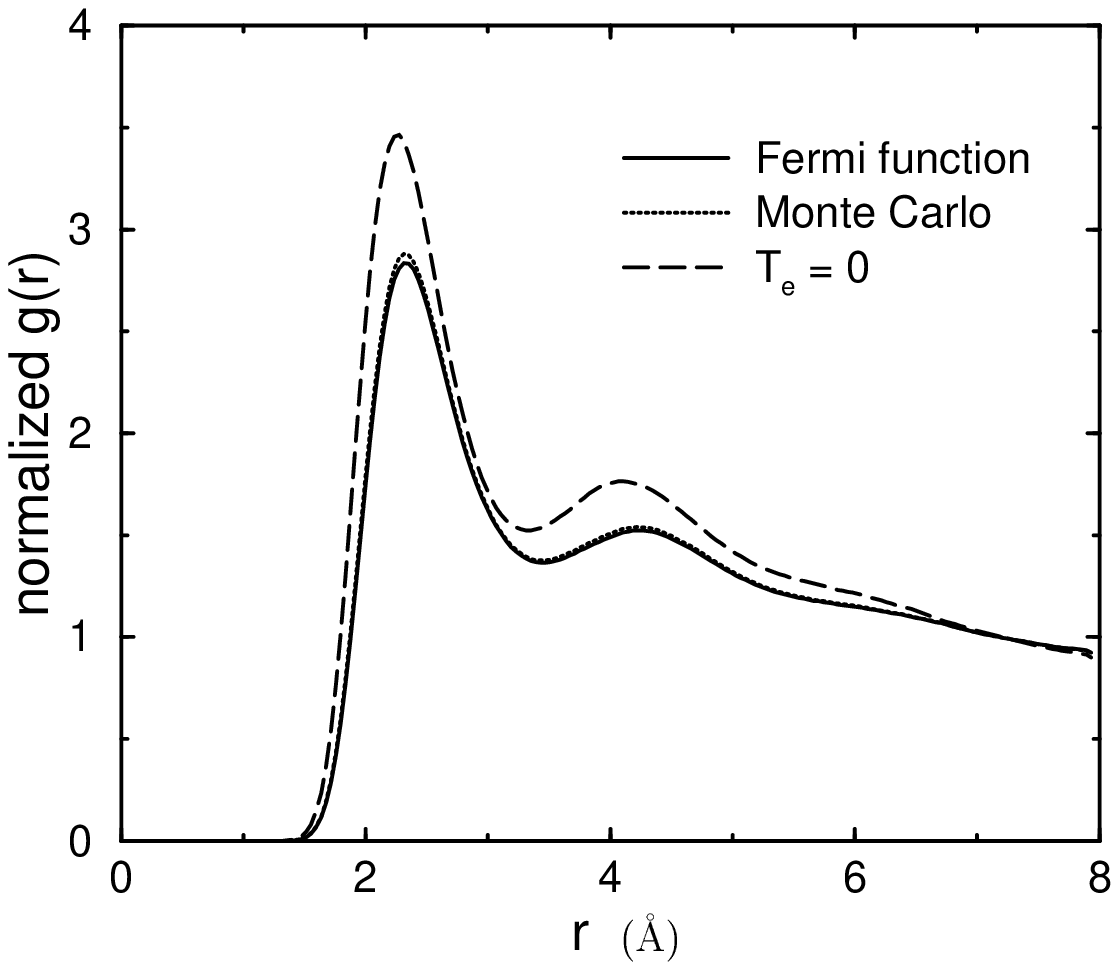}
\vfill}
\caption{Normalized radial density of liquid Ag at density
$\rho=10.5~10^3$~kg.m$^{-3}$ and vibrational temperature $T=5000$~K.
$T_e$ is set either to $T$ or to 0.}
\label{fig:ag}
\end{figure}

We test this algorithm on a model liquid silver, described by a TB Hamiltonian
\cite{tbag}. In Fig.~\ref{fig:ag}, we have plotted the average radial density
$g(r)$ at $T=5000$~K, from three different methods, for a system with 108 Ag
atoms at constant density $\rho=10.5~10^3$~kg.m$^{-3}$. The first method only
considers nuclear moves, the electrons being frozen at $T_e=0$. The
second one is the previous MC method with adjacent trajectories at 2500~K
and 7500~K, and $p=0.1$.
Finally, we perform a single trajectory MC calculation using
only nuclear moves, but fractional occupation numbers given by the Fermi-Dirac
distribution at $T_e=T = 5000$~K. For each atomic
configuration, the chemical potential which normalizes ${\cal N}$
is found by a Newton-Raphson minimization of the error
function $\chi^2(\mu) = [\sum_k n_k^{\rm FD}(\mu)-{\cal N}]^2$ with $n_k^{\rm
FD}(\mu) = [1+\exp(\beta(\varepsilon_k-\mu))]^{-1}$, starting from the
distribution ${\bf N}$ of the previous MC step.
In this case, an entropic correction ($-TS$)
to the energy is included with the usual form given by Mermin \cite{mermin}:
\begin{equation}
S= -k_{\rm B} \sum_k [n_k\ln n_k + (1-n_k)\ln (1-n_k)].
\label{eq:s}
\end{equation}
All simulations consist of $10^7$ nuclear MC cycles. In the second method,
${\cal N}$ electronic moves are attempted for each vibrational move.

The good agreement between the two calculations with $T_e \neq 0$
shows that the present MC algorithm is able to equilibrate both the
electronic and nuclear degrees of freedom, for a quite large system
whose electronic statistical distribution can be safely represented by a
Fermi-Dirac distribution at the same temperature. 

We now turn to finite atomic metal clusters. Finding the electronic average
occupation numbers for any given nuclear geometry is a combinatorial task,
which cannot be solved exactly in the canonical ensemble, except for very few
$(\lesssim 20)$ levels. In the bulk limit, the grand-canonical ensemble is
relevant and gives the FD distribution. For finite,
intermediate sizes, the previous MC algorithm can solve this problem
numerically. However, as is well known for discrete spin systems, exchange
``flip'' moves between occupied and unoccupied neighboring states can be
expected to be mostly rejected at low temperatures $T<\Delta E/k_{\rm B}$,
where $\Delta E = E_{\rm LUMO} - E_{\rm HOMO}$ is the energy gap between the
lowest unoccupied and highest occupied orbitals. The convergence can be greatly
accelerated using the kinetic Monte Carlo (KMC) method \cite{bkl} of Bortz,
Kalos and Lebowitz. We use the KMC
method in conjunction with the local exchange moves, starting from the $T_e=0$
electronic distribution where only the lowest levels are populated.
Tests on small clusters (up to 16 atoms) have shown that about $10^5$ KMC steps
are necessary to ensure convergence towards the exact statistical
\begin{figure}[htb]
\vbox to 6cm{
\includegraphics{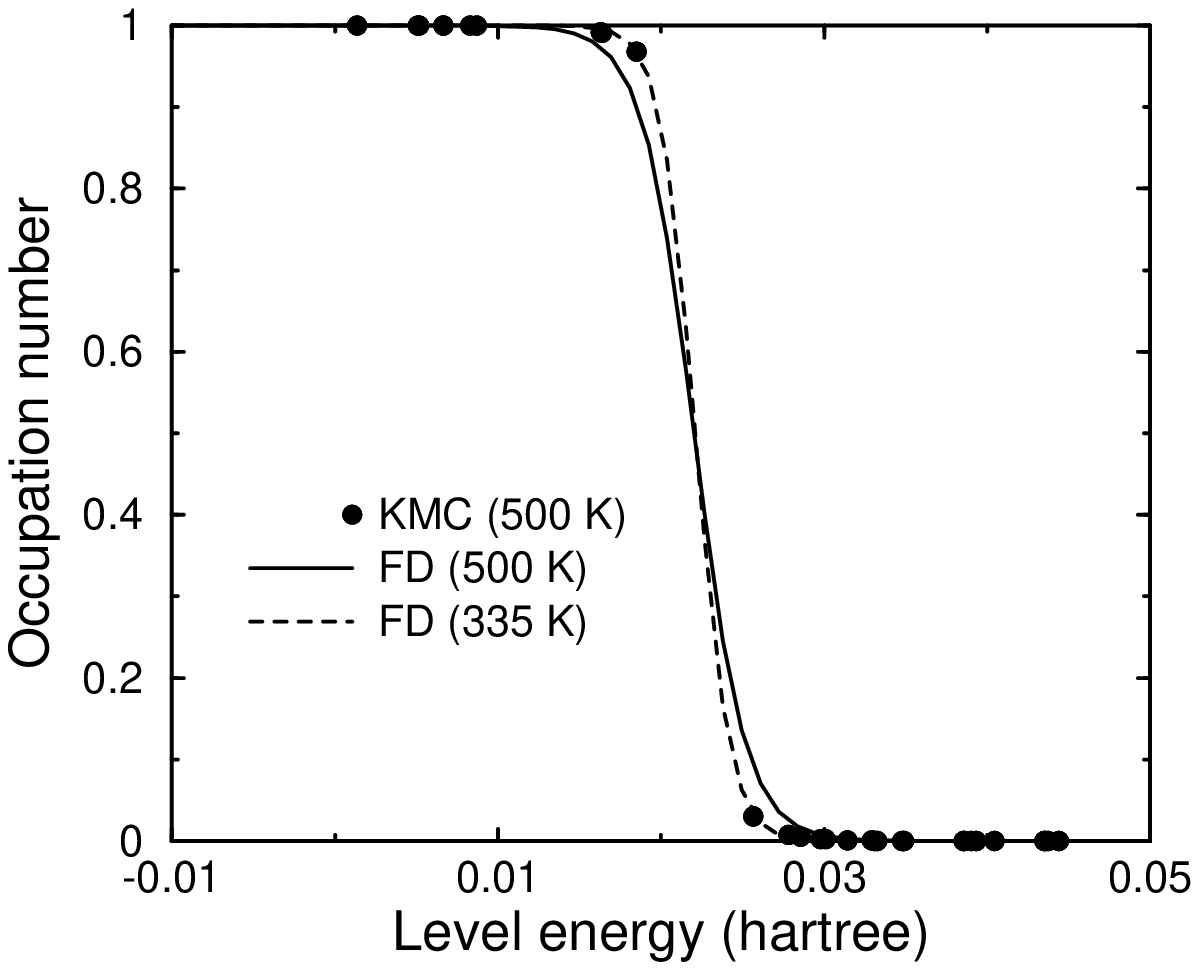}
\vfill}
\caption{Average occupation numbers for the electronic levels of Na$_{40}$
at $T=500$~K. The KMC results are compared with the FD statistics, for which
the best fit is found at 335~K.}
\label{fig:testkmc}
\end{figure}
population. In Fig.~\ref{fig:testkmc} we plot the average occupation
number at $T=500$~K versus level energy for the ground state geometry of
Na$_{40}$ described by a TB Hamiltonian \cite{tbna}.
While the general shape is that of a Fermi-Dirac type, the actual
FD distribution at electronic temperature $T_e=500$~K does not match
the result of the KMC calculation, which is best fitted by a FD law at
effective temperature $T_e^{\rm eff}=335$~K. 

We can combine the KMC algorithm for the electrons with the
usual MC moves for the nuclei. To save computational time, the
electronic problem is solved periodically, once every $M$ steps.
Again, the EMC strategy is used to improve global
convergence and reduce quasi-ergodicity. However, one must be careful when
attempting exchange moves between trajectories at different temperatures,
because the energy depends explicitely on the nuclear coordinates, but also on
temperature via the average occupation numbers. The same problem would hold for
any other temperature-dependent potential, such as the effective potentials
with quantum corrections used in liquids theory \cite{jph}.

More precisely, the acceptance probability of an exchange between
configurations ${\bf R}_i$ and ${\bf R}_j$ initially at the respective inverse
temperatures $\beta_i$ and $\beta_j$ is now
\begin{equation}
{\rm acc}({\bf R}_i \rightleftharpoons {\bf R}_j) = \min[ 1,\exp(\beta_j \Delta
E_j + \beta_i \Delta E_i)],
\label{eq:ptmc2}
\end{equation}
where $\Delta E_k = E({\bf R}_j,\beta_k) - E({\bf R}_i,\beta_k)$, $k=i$ or $j$.
When
the energies are temperature-independent, Eq.~(\ref{eq:ptmc1}) is recovered.
However, in the present case, they include the entropic correction of
Eq.~(\ref{eq:s}) and should be considered as free energies. The canonical
equilibrium of both nuclear and electronic coordinates can be simulated using
this MC method. Since we are dealing with temperature-dependent
energies, useful analysis techniques such as the histograms methods
\cite{histo} cannot be applied here. In addition, corrective terms to the
thermodynamic properties appear due to this dependence. Unfortunately,
because the occupation numbers are obtained numerically, their
derivatives with respect to $\beta$ are hard to get. We can reasonably assume
that the corresponding effects are small at low temperature, since in bulk
metals the electronic heat capacity is only a small quantity with a weak
(linear) dependence upon temperature \cite{solid}.

The previously described MC method is used to simulate the
solidlike-liquidlike phase change in small sodium clusters. Recent theoretical
works \cite{calspi1} emphasized the significant role of geometric
(nuclear) effects on the caloric curves. However, the possible effects of
electrons thermalization have not been considered yet. The influence of
electronic temperature on shape was discussed by Yannouleas and Landman
\cite{landman}, but the jellium model used by these authors does not provide
information about phase changes within an atomistic description. From the
experimental point of view, the results by Haberland and coworkers \cite{hab}
are still far from a complete understanding, especially the complex
size-dependence of the thermal properties.

\begin{figure}[htb]
\vbox to 6cm{
\includegraphics{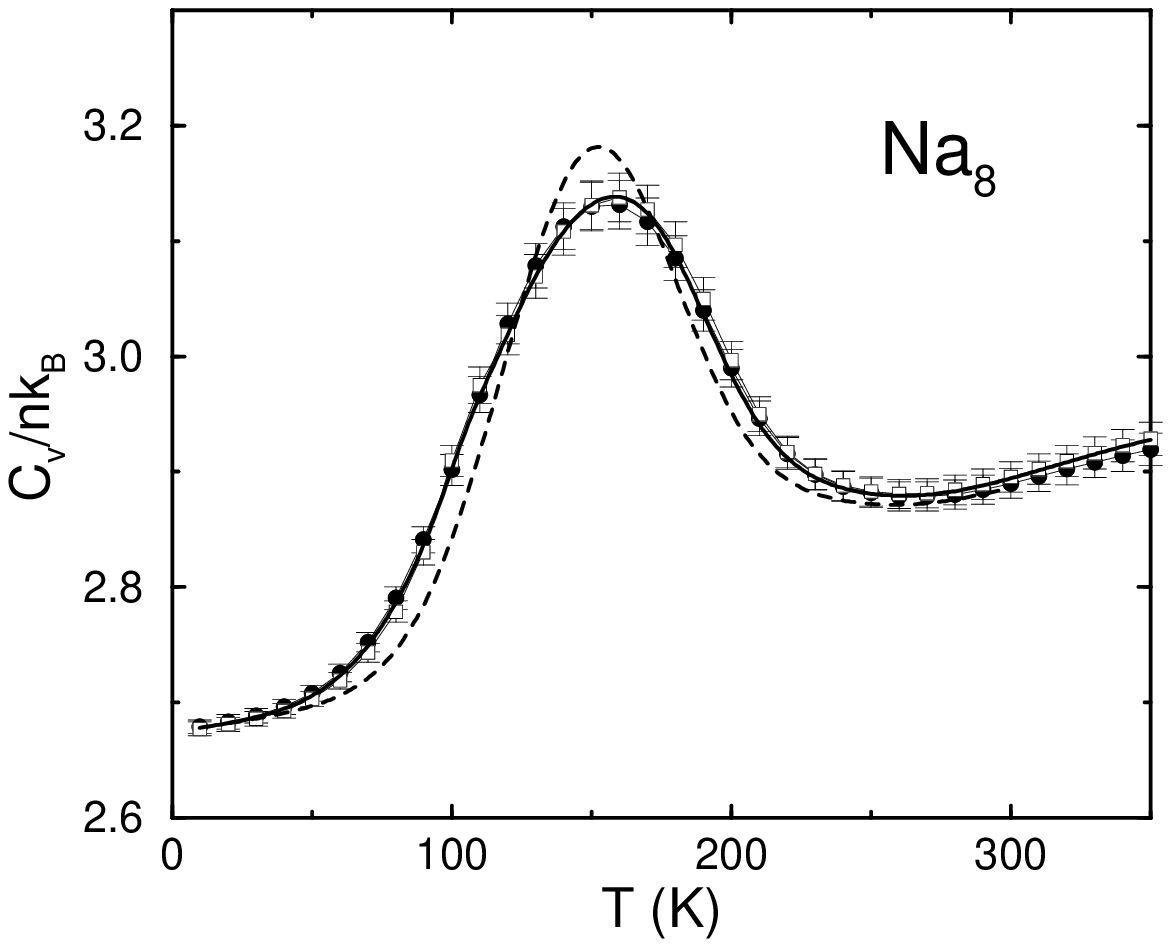}
\vfill}
\caption{Heat capacity per atom of Na$_8$, obtained from exchange Monte Carlo
simulations with frozen (solid line) or thermalized (symbols) electrons.
The average occupation numbers (full circles) are compared with the exact
values (empty squares). The results of simple Monte Carlo of
Ref.~\protect\onlinecite{calspi2} are also shown for comparison (dashed
lines).}
\label{fig:na8}
\end{figure}
We first compare in Fig.~\ref{fig:na8}
the heat capacity curves of Na$_8$ obtained without
consideration of electron temperature, nor use of sophisticated sampling
method \cite{calspi2}, with the ones obtained with the present Monte Carlo
algorithms. Here we can compute the exact
average occupation numbers by solving the combinatorial problem for each
configuration, or we can employ the numerical KMC procedure, both within the
EMC scheme. 35 simultaneous trajectories were propagated with
$10^7$ cycles each, the occupation numbers being calculated for each
configuration.
For comparison, the curves obtained at $T_e=0$ but with EMC moves are
also plotted on Fig.~\ref{fig:na8}. The resulting
heat capacities show a very good agreement between the two present simulations
with nonzero electronic temperature. Thus the KMC scheme provides a good
approach to electronic thermalization. The thermodynamic curves obtained
without using exchange MC, or assuming frozen electrons, are very
similar. Therefore, in this case, we can conclude that (i) the
electronic temperature plays only a small role; and (ii) EMC
does not bring such an improvement over conventional Monte Carlo.

This situation becomes somewhat different for larger sizes.
Na$_{20}$, Na$_{40}$, Na$_{59}^+$ and Na$_{93}^+$ are studied using the same
\begin{figure}[htb]
\vbox to 9.4cm{
\includegraphics{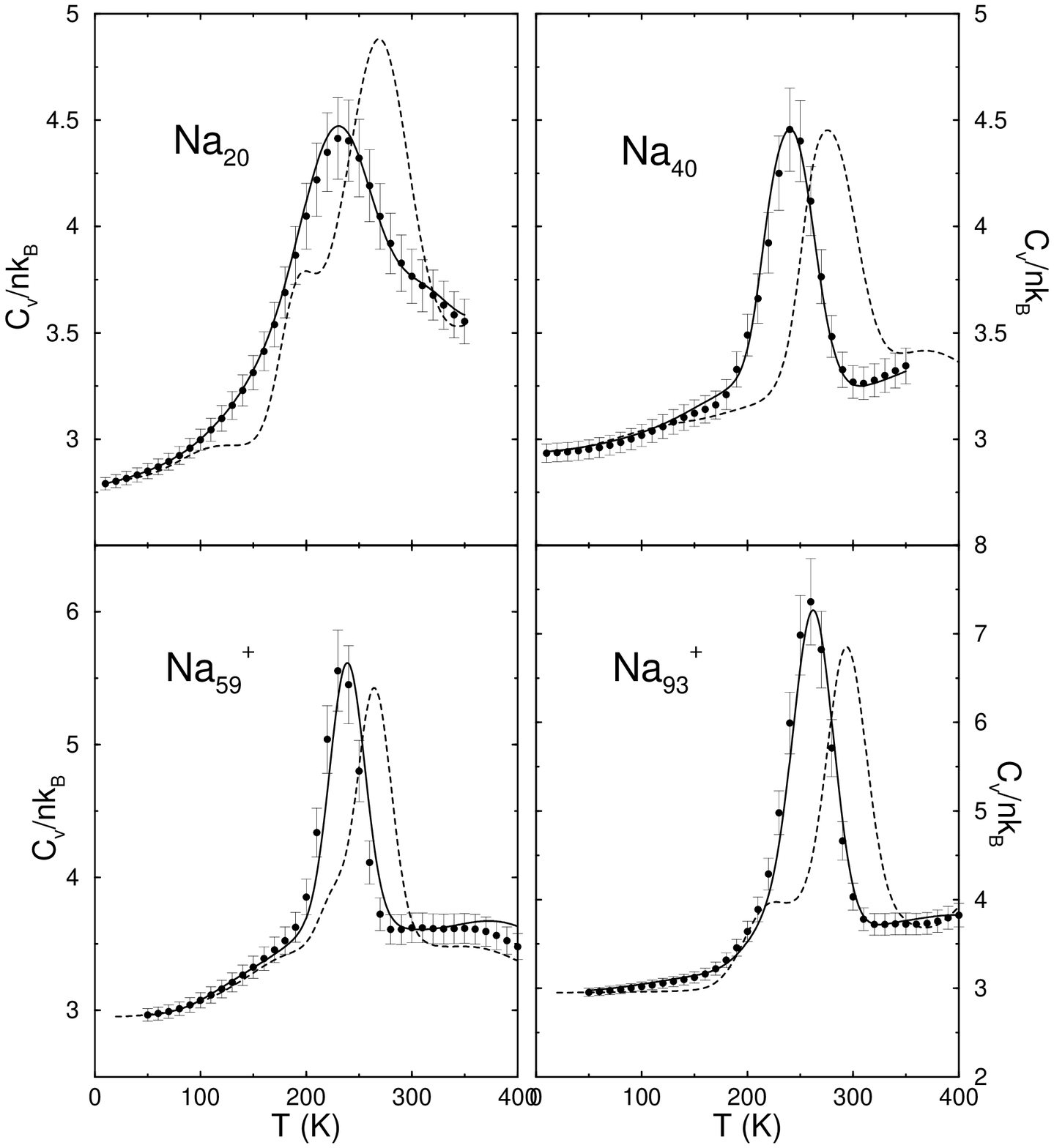}
\vfill}
\caption{Heat capacities of sodium clusters calculated from exchange Monte
Carlo simulations with frozen (solid lines) or thermalized (circles) electrons.
Also shown for comparison are the results of simple Monte Carlo of
Ref.~\protect\onlinecite{calspi2} (dashed lines).}
\label{fig:cvna}
\end{figure}
methods except the one involving the exact calculation of the fractional
numbers. Again, we also compare the heat capacities with our previous
calculations. The whole results are shown in
Fig.~\ref{fig:cvna}. The overall behavior resembles much that of
Ref.~\onlinecite{calspi2}, with a major peak in the heat capacity which marks
the onset of the solidlike-liquidlike phase change. However, two significant
differences can be noted. First, the melting peak appears rather clearly,
without any strong premelting feature (shoulder or peak) at low temperature.
This must be contrasted with most other theoretical studies
\cite{calspi1,calspi2,alonsoblundell} which emphasized multistep melting in
sodium clusters described by various models, but is consistent with experiments
\cite{hab}. Second, the melting temperature
indicated by the top of the peak is shifted to lower temperatures by about
10--30~K depending on size. Again this brings the present results closer to
experiments, as our previous ones with the same tight-binding model were
seen to overestimate melting points in these clusters \cite{calspi2}.
Electronic temperature does not have such a large effect,
since the Monte Carlo results
with $T_e=0$ nearly match the ones with $T_e=T$.
This is not quite surprising, since the temperature required for
electronic transitions is large, even if it decreases as the cluster
grows. However, at least two reasons can be invoked to explain the differences
between the present results and the previously published thermodynamical
curves. Exchange Monte Carlo is known to be a convenient method in
reducing quasi-ergodicity and accelerating convergence \cite{ptmc2}.
Also, by adding the
possibility of changing electronic surface, barrier crossing is further
favored. Thus we expect that the MC methods developed here are
much more efficient than the simple, single trajectories simulations. The new
results are then consistent with a lower melting point \cite{calguet}.

Extensions of the present algorithms to mean-field
monoelectronic Hamiltonians other than
tight-binding is possible. A physical limitation is the relevance of the
calculated excited levels as single-particle states and the neglect of
many-body electron interactions. Another more important limitation
is the difficult combination with molecular dynamics, due to the non-explicit
dependence of the average occupation numbers on the nuclear structure, at least
for finite systems. The use of Car-Parrinello dynamics including entropic
corrections \cite{wentzcovitch} requires some practical
approximations, such as the Fermi-Dirac form for the occupation numbers, or the
assumption that these numbers do not vary much during a short time scale.

The KMC scheme makes the present method naturally suited for
use with Monte Carlo sampling of the nuclear degrees of freedom. Accelerating
procedures \cite{calspi2,mousseau} for the total energycalculations in
tight-binding models can also be a valuable improvement for large
systems. The methods introduced in this Letter have a wide range of
applications, for both finite and infinite systems, metals, insulators or
semiconductor materials. They provide a numerically accurate way of calculating
the fractional occupation numbers, and we gave evidences that the
Fermi-Dirac statistical distribution is not appropriate for small clusters.
Combined with advanced Monte Carlo techniques such as parallel tempering or the
multicanonical ensemble sampling, these methods enable one to investigate the
equilibrium thermodynamics of large complex systems which exhibit various
kinds of phase transitions due to structural isomerization or electronic
excitations. In this respect, clusters close to the insulator-metal crossover
or having magnetic properties offer good candidates for further investigations.


\end{document}